\documentclass[AMA,STIX1COL]{WileyNJD-v2}
\usepackage{moreverb}
\usepackage[utf8x]{inputenc}
\usepackage[T1]{fontenc}

\usepackage{color}
\usepackage{amsmath,amsfonts,amsthm}
\usepackage{pifont}% For checkmarks and crosses
\newcommand{\cmark}{\ding{51}}%
\newcommand{\xmark}{\ding{55}}%
\usepackage{graphicx}
\usepackage{multirow}
\usepackage{textcomp}
\usepackage{listings}
\usepackage{xcolor}
\usepackage{xspace}
\usepackage{url}
\usepackage{hyperref}
\usepackage[capitalise]{cleveref}
\usepackage{pgffor}
\usepackage{algorithm}
\usepackage{balance}
\usepackage{algpseudocode}
\usepackage[caption=false]{subfig}
\usepackage{wrapfig} % For wrapping text around figures

\algnewcommand\algorithmicforeach{\textbf{for each}}
\algdef{S}[FOR]{ForEach}[1]{\algorithmicforeach\ #1\ \algorithmicdo}

\usepackage{todonotes}

\newboolean{showcomments}
\setboolean{showcomments}{true} % toggle to show or hide comments
\ifthenelse{\boolean{showcomments}}
{\newcommand{\nb}[2]{
  \fcolorbox{black}{yellow}{\bfseries\sffamily\scriptsize#1}
  {\sf\small$\blacktriangleright$\textit{#2}$\blacktriangleleft$}
 }
 
}
{\newcommand{\nb}[2]{}
 
}

\usepackage{listings, xcolor}

\definecolor{verylightgray}{rgb}{.97,.97,.97}

\lstdefinelanguage{Solidity}{
	keywords=[1]{anonymous, assembly, assert, balance, break, call, callcode, case, catch, class, constant, continue, constructor, contract, debugger, default, delegatecall, delete, do, else, emit, event, experimental, export, external, false, finally, for, function, gas, if, implements, import, in, indexed, instanceof, interface, internal, is, length, library, log0, log1, log2, log3, log4, memory, modifier, new, payable, pragma, private, protected, public, pure, push, require, return, returns, revert, selfdestruct, send, solidity, storage, struct, suicide, super, switch, then, this, throw, transfer, true, try, typeof, using, value, view, while, with, addmod, ecrecover, keccak256, mulmod, ripemd160, sha256, sha3}, % generic keywords including crypto operations
	keywordstyle=[1]\color{blue}\bfseries,
	keywords=[2]{address, bool, byte, bytes, bytes1, bytes2, bytes3, bytes4, bytes5, bytes6, bytes7, bytes8, bytes9, bytes10, bytes11, bytes12, bytes13, bytes14, bytes15, bytes16, bytes17, bytes18, bytes19, bytes20, bytes21, bytes22, bytes23, bytes24, bytes25, bytes26, bytes27, bytes28, bytes29, bytes30, bytes31, bytes32, enum, int, int8, int16, int24, int32, int40, int48, int56, int64, int72, int80, int88, int96, int104, int112, int120, int128, int136, int144, int152, int160, int168, int176, int184, int192, int200, int208, int216, int224, int232, int240, int248, int256, mapping, string, uint, uint8, uint16, uint24, uint32, uint40, uint48, uint56, uint64, uint72, uint80, uint88, uint96, uint104, uint112, uint120, uint128, uint136, uint144, uint152, uint160, uint168, uint176, uint184, uint192, uint200, uint208, uint216, uint224, uint232, uint240, uint248, uint256, var, void, ether, finney, szabo, wei, days, hours, minutes, seconds, weeks, years},	% types; money and time units
	keywordstyle=[2]\color{teal}\bfseries,
	keywords=[3]{block, blockhash, coinbase, difficulty, gaslimit, number, timestamp, msg, data, gas, sender, sig, value, now, tx, gasprice, origin},	% environment variables
	keywordstyle=[3]\color{violet}\bfseries,
	identifierstyle=\color{black},
	sensitive=false,
	comment=[l]{//},
	morecomment=[s]{/*}{*/},
	commentstyle=\color{gray}\ttfamily,
	stringstyle=\color{red}\ttfamily,
	morestring=[b]',
	morestring=[b]"
}

\newcommand{\ie}{i.e.,\xspace}
\newcommand{\eg}{e.g.,\xspace}

\theoremstyle{definition}

\newtheorem{dfn}{Definition}

\newcommand\BibTeX{{\rmfamily B\kern-.05em \textsc{i\kern-.025em b}\kern-.08em
T\kern-.1667em\lower.7ex\hbox{E}\kern-.125emX}}

\articletype{Research Article}%

\received{<day> <Month>, <year>}
\revised{<day> <Month>, <year>}
\accepted{<day> <Month>, <year>}

\begin{document}

\title{Automated Test-Case Generation for Solidity Smart Contracts: the AGSolT Framework and its Evaluation}

\author[1]{Stefan W. Driessen*}

\author[1]{Dario Di Nucci}

\author[2]{Geert Monsieur}

\author[2]{Damian A. Tamburri}

\author[1]{Willem-Jan van den Heuvel}

\authormark{S.W. Driessen \textsc{et al}}

\address[1]{\orgdiv{Jheronimus Academy of Data Science}, \orgname{Tilburg University}, \orgaddress{\state{Noord-Brabant}, \country{the Netherlands}}}

\address[2]{\orgdiv{Jheronimus Academy of Data Science}, \orgname{Eindhoven University of Technology}, \orgaddress{\state{Noord-Brabant}, \country{the Netherlands}}}

\corres{*Stefan W. Driessen, Sint Janssingel 92, 5211 DA 's-Hertogenbosch \email{s.w.driessen@jads.nl}}

\presentaddress{Sint Janssingel 92, 5211 DA 's-Hertogenbosch}

\abstract[Abstract]{Blockchain and smart contract technology are novel approaches to data and code management that facilitate trusted computing by allowing for development in a distributed and decentralized manner. Testing smart contracts comes with its own set of challenges which have not yet been fully identified and explored. Although existing tools can identify and discover known vulnerabilities and their interactions on the Ethereum blockchain through random search or symbolic execution, these tools generally do not produce test suites suitable for human oracles. In this paper, we present AGSOLT (Automated Generator of Solidity Test Suites). We demonstrate its efficiency by implementing two search algorithms to automatically generate test suites for stand-alone Solidity smart contracts, taking into account some of the blockchain-specific challenges. To test AGSOLT, we compared a random search algorithm and a genetic algorithm on a set of 36 real-world smart contracts. We found that AGSOLT is capable of achieving high branch coverage
with both approaches and even discovered some errors in some of the most popular Solidity smart contracts on Github.}


% \jnlcitation{\cname{%
% \author{S.W. Driessen}, 
% \author{D. Di Nucci}, 
% \author{G. Monsieur}, 
% \author{D.A. Tamburri}, and 
% \author{W. van den Heuvel}} (\cyear{2021}), 
% \ctitle{Automated Test-Case Generation for Solidity Smart Contracts: the AGSolT Framework and its Evaluation}, \cjournal{STVR}, \cvol{2021;00:1--6}.}

\maketitle

\section{Introduction}
\label{sec:introduction}
Blockchain and smart contract technologies are novel approaches to data and code management. They facilitate trusted computing by gracefully allowing for development in a distributed and decentralized manner. Smart Contracts are capsules of code, similar to classes in object-oriented programming languages, such as Java and Python, which are deployed on distributed systems such as blockchains. 
Smart Contracts and blockchains have seen a major rise in popularity in recent years~\cite{anderson2016,luu2016}. In large part, this is due to the inherent qualities of blockchains, such as immutability of data and ease of access to the data stored, which renders extensive testing of critical importance, especially before code deployment. So far, research on testing smart contracts has focused primarily on identifying smart contract- and blockchain vulnerabilities ~\cite{anderson2016,luu2016,delmolino2015,atzei2017}, and applying basic techniques such as fuzzing combined with automated oracles to detect these vulnerabilities~\cite{Jiang2018}\footnote{See also Solfuzzer at \url{https://solidity.readthedocs.io/en/develop/contributing.htm\#running-the-fuzzer-via-afl}}. Automatically detecting vulnerabilities can be a useful tool for smart contract developers, but often having access to a good test suite can prove even more useful to the developer as argued below.

Previous studies have shown that the lack of such test suites is one of the major challenges hampering a successful technology transfer from academics to industry~\cite{Daka2015, Grano2018, Almasi2017}. However, as shown in the extensive investigation conducted by Zou \etal~\cite{Zou2019}, creating test suites for smart contracts is not trivial, and several challenges arise during their implementation. First of all, 54.7\% of the interviewed developers report the lack of powerful tools, including testing tools, for blockchain-specific development. Moreover, no mature testing framework and practical testing guidelines are available. Previous tools such as Oyente~\cite{luu2016} and ContractFuzzer~\cite{Jiang2018} are undoubtedly promising, but they either do not produce test suites at all or very large test suites, which are not human-readable. Furthermore, these tools do not consider all corner cases and scenarios, which is the most critical challenge raised by the developers interviewed in Zou \etal~\cite{Zou2019}. Finally, they note that currently, no tool is available to measure test suite quality of smart contracts. The only potential exception is represented by the tool developed in Wang \etal~\cite{Wang2019b}, who follow a similar approach to ours. However, on the one hand, this tool is not publicly available. On the other hand, it does not consider test suite size explicitly, which might result in unnecessarily long and complex test suites.\\

In this paper, first, we analyze some of the properties for designing a tool for automated test case generation for smart contracts. We believe that these properties are also relevant to researchers who focus on bug detection with automated oracles. We find that properties previously identified~\cite{7102580} for popular programming languages such as Java and C still hold (\eg deciding on quality metrics and covering corner cases), but additional qualities are desirable, which we describe below.
Then, to handle these challenges mentioned above, we present \textsc{AGSolT}\footnote{\url{https://github.com/AGSolT/AGSolT2021Submission}\label{ftn:agsolt_appendix}} (Automated Generator of Solidity Test Suites), an automated test case generation tool for unit testing for the smart contract programming language Solidity on the Ethereum blockchain. AGSolT creates concise test suites for individual smart contracts while aiming to achieve a high branch coverage level. Existing literature on reducing the size of test suites and test cases can be considered a good initial step for making more human-readable test suites, which are more likely to be used in practice 
~\cite{panichella2018, Fraser2013}. Therefore, \textsc{AGSolT} aims at creating smaller test suites, making unit testing\footnote{In the domain of smart contracts, \textit{unit testing} means testing individual smart contracts.} and regression testing easier~\cite{Shamshiri2015,Yoo2010}. \textsc{AGSolT} could lead the way to create higher-quality test suites for Solidity smart contracts that exercise more in-depth corner cases and scenarios combining metaheuristic techniques for the automated test-case generation with developers-provided oracles.

We equip \textsc{AGSolT} with two common approaches for automated test case generation: (1) fuzzing, which is a random testing approach, and (2) genetic algorithms, which are a search-based testing approach. On the one hand, fuzzing generates test cases randomly; on the other hand, the genetic algorithms iteratively improve an initially random set of test cases through a search guided by one or more objective functions. Previous research~\cite{Arcuri2010, Shamshiri2018} has shown that both approaches can be equally effective when generating test suites, which makes them both valid approaches to an automated test case generation problem.

We conducted an empirical study on 36 real-world smart contracts to assess the effectiveness of \textsc{AGSolT} and take a closer look at how the two approaches compare for Solidity smart contracts. As far as the authors are aware, this is the first comparison of the sort in the domain of smart contracts.
We find that \textsc{AGSolT} achieves good branch coverage on a variety of smart contracts and can detect errors in some of the most popular smart contracts on Github. Both approaches show promise for future investigation, although genetic algorithms might be slightly more suitable for achieving branch coverage on specific types of smart contracts. Although neither approach is significantly faster than the other, our experiments seem to indicate that a guided search that prefers smaller test cases might be better at reducing the time spent running the tests on a blockchain implementation.

In sum, this paper contributes to the state-of-the-art by:

\begin{enumerate}
    \item Proposing a set of challenges specific to the blockchain domain that any automated test case generation tool should aim to overcome.
    \item Introducing \textsc{AGSolT}: an automated test case generation tool, capable of: (i) creating small, human-readable test suites that are optimized for branch coverage; (ii) allowing for the implementation of different types of algorithms such as random-testing and search-based-testing; (iii) being easily adapted to allow for different types of objectives such as mutation coverage or statement coverage.
    \item Providing the first comparison between a guided search and a random search in the domain of automated test case generation for smart contracts.
\end{enumerate}

The rest of this paper is organized as follows: \cref{sec:background} introduces the concept of smart contracts in the context of the Ethereum blockchain and discusses existing ATG tools for these smart contracts.
\Cref{sec:problem_statement} formalizes the challenges that we identify for creating an ATG tool for Smart contracts.
In \cref{sec:AGSolT}, the \textsc{AGSolT} tool is introduced, and its workings are explained.
\Cref{sec:experimental_evaluation} describes the design and the results of the empirical study we conducted to evaluate \textsc{AGSolT} and compare the search-based and random algorithms, while \cref{sec:threats} discusses its threats to validity.
Finally, \cref{sec:conclusion} discusses the results of these experiments and introduces potential future work.

\section{Background}
\label{sec:background}
This section provides an overview concerning blockchain, smart contracts, and their testing.

\subsection{The Ethereum Blockchain and Smart Contracts}
\label{sec:smart_contracts}
A blockchain~\cite{nakamoto2008,buterin2013} can be viewed as a \textit{decentralized}, \textit{distributed} digital ledger: an ordered list of \textit{blocks}, which themselves contain an ordered list of \textit{transactions}.
New blocks are added by \textit{miners}, who follow a \textit{consensus protocol} that dictates the rules of the blockchain, including how to add new data and deal with conflicting versions of the blockchain.
On the Ethereum blockchain, transactions can be used to transfer \textit{Ether} (ETH) cryptocurrency from one address to another and deploy and interact with \textit{smart contracts}. Ether is also used to compensate the miner, who receives a small fee (called \textit{Gas}) from the transaction sender for registering a transaction on the blockchain. In addition to sending Ether, transactions can also be used to deploy- and interact with smart contracts
Because of its inner working, the Ethereum blockchain can store (almost) any data type, so long as modifications are made in a transaction-based manner.
Its creators have leveraged this property to store (compiled) pieces of code, called smart contracts, on the blockchain, which can be used as follows.
Each transaction has a ``Data'' field where a transaction sender can store bytecode.
When sending to a previously unused \textit{address}, this bytecode can be interpreted by miners that use the \textit{Ethereum Virtual Machine} (EVM) to create new smart contracts whose bytecode is stored on the blockchain at the new address.

After a contract has been deployed, the transactions sent to its address can invoke the execution of the code stored on the blockchain by including the method to be invoked and any input parameters in the Data field of the transaction. The EVM specifies how to alter the state of the system based on the Data field of the transaction and the code stored at the specified address~\cite{wood2014}.
If a transaction is issued without a recognizable method in its ``Data'' field, a special function called \textsc{fallback} is invoked.

Since writing bytecode by hand is impractical, several high-level programming languages have been created, the most popular of which is \textit{Solidity}, which is inspired by Python, C++, and Javascript~\footnote{\url{https://solidity.readthedocs.io/en/v0.5.8/}}.
Smart contracts in Solidity are similar to \textit{classes} in object-oriented programming and behave similarly to \textit{objects}: the smart contract code serves as a blueprint to deploy many instances on the blockchain, each with their address and internal state. Similarly, Solidity smart contracts have both public functions and variables that can be accessed from outside the smart contract and private functions and variables that can only be interacted with by the contract itself.

\subsection{Smart Contract Weaknesses and Testing}
Detecting vulnerabilities in smart contracts has been a hot research topic in recent years, especially since the infamous DAO (Distributed Autonomous Organisation) attack in 2016, where roughly 60 million dollars worth of Ether was stolen because of an unforeseen exploit in a published smart contract~\cite{castillo2016}.
Due to specific blockchain properties, such as the immutability of committed blocks and its distributed and decentralized nature\footnote{Distributed in this context implies that anyone can access the bytecode of a smart contract, decentralized means that anyone can interact with a deployed contract.}, the proper implementation of smart contracts is particularly challenging.
We discuss the most relevant literature below to illustrate this point.

Delmolino \etal~\cite{delmolino2015} found that when teaching undergraduate students to create smart contracts, even simple implementations lead to a multitude of non-trivial problems.
Often, such problems do not prevent compilation but leave the contract vulnerable to exploitation or unintended behaviors.
Anderson \etal~\cite{anderson2016}, Luu \etal~\cite{luu2016}, and Atzei \etal~\cite{atzei2017} investigated already published contracts and highlighted that some of them present design flaws although already published on the blockchain.
Recently, Zou \etal~\cite{8847638} investigated the challenges related to smart contract testing and confirmed that almost half of all developers desired tools to verify code correctness.
The above studies and the previously mentioned DAO attack motivated introducing new development tools to develop and test safe smart contracts effectively\footnote{ \url{https://github.com/ethereum/wiki/wiki/Safety\#ethereum-contract-security-techniques-and-tips}}.
% For example, the popular smart contract developing platform \textit{Truffle}\footnote{Available online at \url{https://truffleframework.com/}} has adopted and integrated \textit{Mocha}\footnote{Available online at \url{https://mochajs.org/}}, a JavaScript testing framework, and \textit{Chai}\footnote{Available online at \url{https://www.chaijs.com/}} a collaborating assertion library to work with Solidity, making it more intuitive for developers to write unit tests for their smart contracts~\cite{mocha_for_truffle}.
Several tools have been put forward that we introduce briefly below:
\textsc{Solidity-coverage}~\footnote{\url{https://blog.colony.io/code-coverage-for-solidity-eecfa88668c2/}\label{solidity-coverage}} measures the quality of an existing test suite by checking whether \textit{branch coverage}~\cite{Anand2013} has been achieved (\ie whether all possible paths through the code have been executed).
\textsc{SolidityCheck}~\cite{zhang2019soliditycheck} checks Solidity code for patterns that are known to lead to vulnerabilities and warns the user about them.
Wu \etal~\cite{wu2019mutation} have designed 15 mutation operators for Solidity smart contracts and use these to detect defects in 26 real-world smart contracts.
\textsc{Oyente}~\cite{luu2016} creates a control-flow graph for a given smart contract and uses symbolic execution to check its \textit{branch feasibility}, (\ie whether each part of the code is theoretically reachable), as well as whether vulnerabilities are present.
% \textsc{SmartShield}~\cite{9054825} is a bytecode rectification tool that automatically fixes three types of security bugs.
\textsc{ADF-GA}~\cite{Zhang2020} uses control-flow-graphs with dup-based covering criteria but only tests on a small set of smart contracts that only use integers and unsigned integers. Similarly, Wang \etal ~\cite{Wang2019b} propose a tool propose a tool for creating branch-covering test suites that they test on 8 smart contracts. A recent addition by Liu et. al. is \textsc{ModCon}, which relies on user-defined models to impose model testing on smart contract \cite{Liu2020MAM}.
Finally, \textit{fuzzers}~\cite{sutton2007} automatically create test cases for smart contracts by generating random (within a specified range) inputs for contract functions to detect errors.
% \textsc{Solfuzzer}~\cite{solfuzzer} has been created by the Solidity developers to detect internal compilation errors and segmentation faults.
The commercial \textsc{Echidna}~\cite{smith2018} tries to break user-defined invariants, while the academic \textsc{ContractFuzzer}~\cite{Jiang2018} checks for both coding errors and the vulnerabilities mentioned by Luu \etal~\cite{luu2016} and Bartoletti \etal~\cite{bartoletti2017}.

When it comes to automated unit-testing of Solidity smart contracts on the Ethereum blockchain, each of the approaches mentioned above comes with its limitations: (i) \textsc{Solidity-coverage}\textsuperscript{\ref{solidity-coverage}}, \textsc{Solidity Check}~\cite{zhang2019soliditycheck}, and \textsc{Oyente}~\cite{luu2016} do not produce test suites; (ii) \textsc{Echidna}~\cite{smith2018} and \textsc{ModCon}~\cite{Liu2020MAM} require the user to define invariants or models of their code, and (iii) \textsc{ADF-GA}~\cite{Zhang2020} and tool of Wang \etal are tested on small subsets of possible smart contracts and do not provide online code to be used for further research. \textsc{Contractfuzzer}~\cite{Jiang2018} is perhaps the most complete approach out there because it creates test suites fully automatically and works on a variety of smart contracts. However, the tool focuses on vulnerability detection through automated oracles as opposed to creating \textit{test suites} which can be used by human oracles. Additionally, existing literature has suggested that random search approaches (\eg fuzzing) run the serious risk of being too simplistic to fully capture corner cases in more complex applications when compared to guided search approaches~\cite{Shamshiri2018,harman2010}. 

For these reasons, we introduce \textsc{AGSolT} (Automated Generation of Solidity Test Suites). This tool can easily leverage different search algorithms to automatically generate test suites for Solidity smart contracts that aim to achieve branch coverage. In the next sections, we first introduce the challenges that \textit{any} tool or framework that sets out to achieve this goal will meet and then discuss how \textsc{AGSolT} aims to overcome these challenges. Finally, we demonstrate the effectiveness and efficiency of the tool by experimenting on 36 real-world smart contracts.

\section{Smart Contract Testing}
\label{sec:problem_statement}
This section introduces a set of properties that an effective automated test suite generation tool should possess. These properties were found through iterative experimentation, aiming at finding the corner cases not covered by the two search strategies exploited by \textsc{AGSolT}. In particular, we started implementing existing algorithms and looking at the branches that these were unable to cover and identified the causes. We describe here, those blockchain-specific properties we found, which have not yet been identified in the literature. We direct the reader towards existing literature \cite{mitchell1998, Laumanns2002, Deb-02, Scalabrino2016, rojas2017} for more background on the general challenges of test case generation, such as choosing objectives, objective functions and improving efficiency and effectiveness. We divide these properties into three different types: \textit{transactional properties} of blockchains and smart contracts, \textit{properties of the blockchain} on which the smart contract is deployed, and properties that define the way smart contracts \textit{interact} with other smart contracts. We argue that any tool for automated test-case generation should consider these properties.

\subsection{Transaction Properties}
The only way to change the state of a smart contract is by sending a \textit{transaction} to the contracts' address and invoke one of its functions.
Besides the function and parameter specification, \textit{every} interaction with a smart contract has to provide a sender, which is the address from which the transaction was sent, a value\footnote{Many blockchain interaction platforms do not require a value be specified, in which case this defaults to zero.}, which is the amount of Ether sent in the transaction, and an amount of gas, which is the fee that the sender has to pay to the miner for the computational power involved in adding this transaction to the block.
These \textit{transaction properties} can be accessed by the smart contract receiving the transaction and influence its inner workings, affecting which branches are traversed.

\begin{lstlisting}[language=Solidity, caption=An example of Ethereum-specific properties., label=sc:eth_specific]
  pragma solidity 0.5.12;
  
  contract Auction {
    address payable public Seller;
    address payable public Frontrunner;
    uint public HighBid;
    uint public CloseTime;
  
    constructor(uint _CloseTime) payable public {
      Seller = msg.sender;
      Frontrunner = msg.sender;
      HighBid = msg.value;
      CloseTime = _CloseTime;
    }
  
    function Bid() payable external{
      require(msg.value > HighBid);
      Frontrunner.transfer(HighBid);
      HighBid = msg.value;
      Frontrunner = msg.sender;
    }
  
    function Claim() external{
      require(block.timestamp > CloseTime);
      // Implement ownership transfer
      selfdestruct(Seller);
    }}
  \end{lstlisting}  

As an illustration, Smart Contract \ref{sc:eth_specific} shows an example of a simple Auction on the Ethereum blockchain.
When the contract is initiated the constructor is executed, which instantiates the \textsc{Seller}, \textsc{Frontrunner}, \textsc{HighBid} and \textsc{CloseTime} variables. Afterwards anyone can make a Bid by calling the \textsc{Bid()} function (lines 16-21). This function first checks if the transaction property \textsc{msg.value} (the new bid) is higher than the current highest bid and if it is, it refunds the previous highest bidder (\textsc{Frontrunner}) and changes the Highest Bid and frontrunner based on the transaction information \textsc{msg.value} and \textsc{msg.sender} respectively.

Any automated test-case generation tool for smart contracts should generate test-cases containing transactions from different accounts to test sender-dependent functionality.
Similarly, the tool should vary the amount of Ether send with a transaction and evolve it either as if they were input variables or chosen for this purpose.

\subsection{Blockchain Properties}
Besides transaction properties, a smart contract has access to additional information from the blockchain environment on which it is deployed, such as the address of the miner of the current block, the gas limit of the current block (\ie the maximum amount of computation in a block), the hash of any of the least 256 recently added blocks, and the time and block number of the current block. Moreover, because each smart contract has an address, it has a balance in Ether associated with it, which affects its ability to send Ether.
An example of this is given by the \textsc{Claim()} function in Smart Contract \ref{sc:eth_specific} which compares the blockchain property \textsc{block.timestamp} (which gives the time since the Unix epoch for this block) with the user-specified \textsc{CloseTime} before the auction can be closed. If the specified time has been reached, the smart contract removes itself from the blockchain and sends its entire balance to the seller.
These blockchain properties can be manipulated (within certain limitations) by the test environments. A useful test case generation tool should vary some or all of these blockchain properties for better testing while at the same time respecting the logical rules of the blockchain, such as that block numbers and time must always increase between different blocks.

\subsection{Interactive Properties}
Similarly to how Java classes can instantiate and interact with other classes, smart contracts on the Ethereum blockchain can instantiate and send transactions, such as method invocations or Ether transfers, to other smart contracts.
However, there are two essential differences.
First, smart contracts can send transactions to any address on the blockchain, allowing them to transfer Ether to a wallet or call functions of \textit{any} smart contract on the same blockchain as long as that contracts address is passed as a variable to the calling smart contract. A special case occurs when the contract sends a transaction to the so-called \textit{zero-address} (\textsc{0x0}), which causes the instantiation of a new smart contract.
Second, smart contracts have an Ether balance that they can use to send Ether alongside transactions to other smart contracts to invoke the contract's functionality or purely transfer currency.

At line 18 of the \textsc{Bid()} function in Smart Contract \ref{sc:eth_specific} the previous \textsc{Frontrunner} is refunded the bid.
If such \textsc{Frontrunner} is a smart contract, this transaction invokes the fallback function of that smart contract, which could, in turn, call one of the functions in the \textsc{Auction} smart contract.

Automated Test Case Generation Tools for Smart Contracts should be aware of both existing addresses, as well as non-existing addresses and the zero-address, which might cause errors in the smart contract. Additionally, they should anticipate interaction with smart contracts outside the programmer's control.

\section{AGSolT: Automated Generator of Solidity Test Suites}
\label{sec:AGSolT}
\begin{wrapfigure}{r}{0.4\textwidth}
    \centering
    \includegraphics[width=0.38\textwidth]{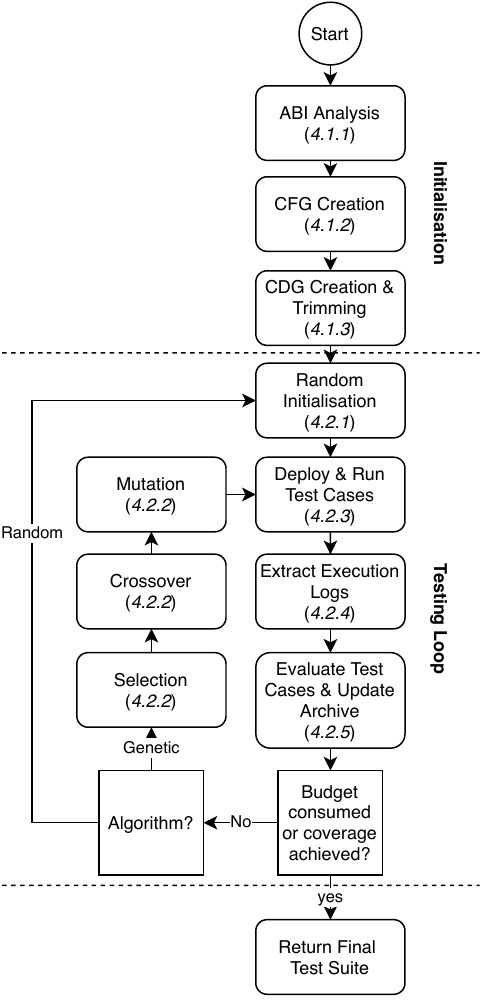}
    \captionsetup{margin=0.025\textwidth}
    \caption{The flowchart of \textsc{AGSolT}; each step is explained in a corresponding section.}
    %\Description[The flowchart of AGSolT.]{The flowchart of AGSolT.}
    \label{fig:Flowchart_AGSolT}
\end{wrapfigure}

This section describes the design choices and algorithmic procedures that make up the main workings of \textsc{AGSolT}.
\Cref{fig:Flowchart_AGSolT} shows its high-level workings, which is composed of an \textit{initialization phase} and a \textit{testing loop}. In the initialization phase, relevant properties of the smart contract(s) under examination are extracted, which are required during the testing loop. During the testing loop, test cases are run on a blockchain implementation\footnote{\url{https://www.trufflesuite.com/ganache}} and their performance is evaluated using the branch distance. \textsc{AGSolT} is mostly implemented in Python, except the instrumentation of the blockchain, which is done through the \textsc{web3}\footnote{\url{https://web3js.readthedocs.io}\label{ftn:web3}} library in Javascript.
Our ultimate goal is to achieve branch coverage. This concept can be intuitively understood by viewing a piece of code as a graph with nodes that contain grouped statements and edges that indicate jumps between these groups brought about by conditions (such as if-else statements) that control which groups of statements are executed and which are not. Sections \ref{sec:cfg_creation} and \ref{sec:cdg_creation_and_optimization} describe how \textsc{AGSolT} creates these graphs and gives some examples for further illustration. In sections \ref{sec:test_cases_and_random_initialisation} and \ref{sec:dynamosa_alg} we demonstrate the two implemented approaches for creating test suites whose test cases traverse all (or as many as possible) of these edges. \textsc{AGSoLT} considers each branch to be covered, as a separate objective and tries to improve upon each objective simultaneously, thus offering multi-objective optimization.

\subsection{Initialization Phase}
During the initialization phase, \textsc{AGSolT} extracts several characteristics of the smart contract under investigation to create the first generation of test cases that can be improved in the \textit{testing loop}.

\subsubsection{ABI Analysis}
\label{sec:ABI_Analysis}
When Solidity code is compiled into bytecode, an \textsc{Application Binary interface} (ABI) file is created.
This file contains the basic information necessary for test case generation (\ie function names and input types). Additionally, during this step, hard-coded values of the contract are scraped to be used for \textit{seeding} the method invocations. Seeding is a common technique used in automated test case generation, which involves including certain values with higher probability when randomly selecting variables for input variables~\cite{fraser2012seed}. In \textsc{AGSolT} whenever a random input variable or ETH value or account is selected, first a check is performed whether one or more hard-coded values of the corresponding type were present in the smart contract (and consequently scraped). If there are such values, 50\% of the time a random scraped value is picked instead of a completely random value. This allows \textsc{AGSolT} to automatically leverage information inside the smart contract to reach branch coverage quicker.

\begin{table}[ht]
    \centering
    \caption{EVM opcodes needed to generate test cases}
    \label{tab:opcodes}
    \begin{tabular}{|c|c|c|c|c|}
        \hline
        Hex & Opcode & Stack Input & Stack Ouput \\
        \hline
        56 & JUMP & dest & \\
        57 & JUMPI & dest, bool & \\
        5B & JUMPDEST & & \\
        10 & LT & $a$, $b$ & $a<b$\\
        11 & GT & $a$, $b$ & $a>b$\\
        12 & SLT & $a$, $b$ & $a<b$\\
        13 & SGT & $a$, $b$ & $a>b$\\
        14 & EQ & $a$, $b$ & $a=b$\\
        15 & ISZERO & $a$ & $a=0$\\
        \hline
    \end{tabular}
\end{table}

\subsubsection{Control Flow Graph Extraction}
\label{sec:cfg_creation}
To keep track of the branches to be traversed and those already covered, \textsc{EvoSol} extracts the \textit{control dependency graph}~\cite{ferrante1987program} of the smart contract.
To this end, first the \textit{Control Flow Graph} (CFG) is distilled from the bytecode using the python \textsc{evm\_cfg\_builder} library\footnote{\url{https://github.com/crytic/evm\_cfg\_builder}}. A conceptual explanation of how a CFG can be extracted from bytecode is given in our online appendix\footnote{\url{https://github.com/AGSolT/AGSolT2021Submission/tree/master/CFG\_Creation\_Appendix}}.
The reason the CFG is created from the bytecode (as opposed to the Solidity code), is because it makes it possible to extract the values that are on the stack when the EVM evaluates a predicate controlling a branch.
These values are needed later for deciding how close a test case is to satisfying a predicate, and consequently, traversing the branch it controls.
\cref{tab:opcodes} shows the 9 opcodes which are relevant for identifying nodes and branches in the opcode column, their hex value as it appears in bytecode as well as the argument(s) they consume from the stack and the output value they push onto the stack. The "JUMP"-opcode is used to jump to a different part of the bytecode for execution (indicated by the destination value). The "JUMPI"-opcode works similar to "JUMP" except that execution continues from the destination, only if the consumed bool is true, this is what creates branches in the CFG. Finally the other opcodes shown in \cref{tab:opcodes} correspond to the predicates that can control a branch; $<$, $>$, $==$ and $\neg$. Note that $\leq$, $\geq$ and $\neq$ can be represented with $\neg>$, $\neg<$ and $\neg==$ respectively.
For each branching node, \textsc{AGSolT} identifies the opcode that corresponds to the controlling predicate to compute the branch distance for the outgoing branches.

\begin{algorithm}
\caption{\textsc{CompactifyCFG}}
\label{alg:compactifyCFG}
\footnotesize
\begin{flushleft}
{\textbf{Input:}}\\
$N$ \Comment{The set of all nodes in the CFG.}\\
{\textbf{Output:}}\\
$N'$ \Comment{The set of nodes where nodes with superfluous branches have been merged.}\\
\end{flushleft}
\begin{algorithmic}[1]
\Procedure{CompactifyCFG}{}
\State $\textit{UN} \xleftarrow{} N$ \Comment{Initialise the unmerged nodes.}
\State $\textit{MN} \xleftarrow{} \emptyset$ \Comment{Initialise the merged nodes.}
\While{$\textit{UN} \neq \emptyset$}
\State $\textit{Node} \xleftarrow{} \textit{any } n\in \textit{UN}\text{ }|\text{ }n.\textit{incoming\_nodes} \cap \textit{UN} = \emptyset$
\State $\textit{UN} \xleftarrow{} \textit{UN} - \textit{Node}$
\State $\textit{MN} \xleftarrow{} \textit{MN} \cup \{\textit{Node}\}$
\State $\textit{UN, MN} \xleftarrow{} {\sc Compactify}(\textit{Node}, \textit{UN}, \textit{MN})$
\EndWhile\\
\Return $\textit{MN}$
\EndProcedure
\end{algorithmic}
\end{algorithm}

\begin{algorithm}
\caption{\textsc{Compactify}}
\label{alg:compactify}
\footnotesize
\begin{flushleft}
\textbf{Input:}\\
$\textit{Node}$ \Comment{A node to be compactified.}\\
$\textit{UN}$ \Comment{The set of unmerged nodes.}\\
$\textit{MN}$ 
\Comment{The set of merged nodes.}\\
\textbf{Output:}\\
$\textit{UN}'$ \Comment{The updated set of unmerged nodes.}\\
$\textit{MN}'$ \Comment{The updated set of merged nodes.}\\
\end{flushleft}
\begin{algorithmic}[1]
\Procedure{Compactify}{}
\If{$\# \textit{Node.outgoing\_nodes} \neq 1$}
\Return \textit{UN, MN}
\ElsIf{$\# \textit{Node.incoming\_nodes} \neq 1$}
\Return \textit{UN, MN}
\Else
\State $\textit{nextNode} \xleftarrow{} \textit{Node.outgoing\_node}$
\State $\textit{UN} \xleftarrow{} \textit{UN} - \textit{nextNode}$
\State $\textit{MN} \xleftarrow{} \textit{MN} \cup \{\textit{nextNode}\}$
\State $\textit{Node} \xleftarrow{} \textit{Node} \bigoplus \textit{nextNode}$
\EndIf\\
\Return $\textsc{Compactify(\textit{Node}, \textit{UN}, \textit{MN})}$
\EndProcedure
\end{algorithmic}
\end{algorithm}

\subsubsection{Control Dependency Graph Creation and Optimization}
\label{sec:cdg_creation_and_optimization}
Some edges of the graph lead to superfluous nodes whose execution neither leads to or depends on any predicate, that could waste part of the search budget.
Therefore, they are eliminated by running the \textsc{CompactifyCFG} algorithm shown in \Cref{alg:compactifyCFG} which uses the \textsc{Compactify} procedure in \Cref{alg:compactify}.
Finally, \textsc{AGSolT} uses the algorithm proposed by Lengauer and Tarjan~\cite{Lengauer1979} to determine the control dependencies between the nodes and distil the \textit{Control Dependency Graph} from the \textit{Control Flow Graph}.
Considering some specific characteristics of the Ethereum blockchain, the graph can still be optimized by removing some nodes that are not relevant for test case generation.
These nodes and edges belong to the following patterns:

\begin{itemize}
    \item \textbf{Dispatcher Nodes and Edges.} The \textsc{Ethereum} bytecode contains a dispatcher function that handles the transactions to the smart contract.
    Since \textsc{AGSolT} invokes all (public) methods, there is no need to calculate the branch distance for these edges.
    \item \textbf{Empty Fallback.} An empty fallback function is initialized when the user does not explicitly define one.
    However, such a function can be safely ignored as it does not change the semantics.
    \item \textbf{State Variables.} Public variables are accessed as functions through the contract dispatcher.
    Since calling these variables does not help cover new branches, the corresponding nodes and edges in the CDG can be ignored.
\end{itemize}
    
\begin{wrapfigure}{r}{0.4\textwidth}
    \centering
    \captionsetup{margin=0.025\textwidth}
    \includegraphics[width=0.35\textwidth]{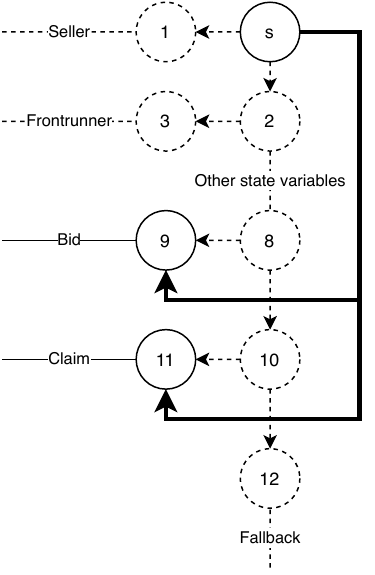}
    \caption{CDG of the {\tt dispatcher} function in Smart Contract. \ref{sc:eth_specific}}
    %\Description[CDG of the {\tt Claim} function in Smart Contract \ref{sc:eth_specific}]{The superfluous nodes in the dispatcher, state variables and fallback of Smart Contract \ref{sc:eth_specific}.}
    \label{fig:contract_dispatcher}
\end{wrapfigure}

An example of these patterns is shown in \cref{fig:contract_dispatcher}: The control dependency graph of smart contract \ref{sc:eth_specific} starts with dispatcher nodes (even nodes), which are used to identify the method or state variable (starting at uneven nodes) that is called. If none of the state variables or methods was passed in the transaction, the fallback function (starting at node 12) is invoked. Since no fallback function was specified in smart contract \ref{sc:eth_specific}, this method is empty, and neither invoking it nor any of the state variables is particularly interesting for testing purposes. For that reason \textsc{AGSolT}, removes the edges and nodes corresponding to the dispatcher, state variables and empty fallback functions (shown dotted in \cref{fig:contract_dispatcher} and creates new edges to the relevant methods (\textsc{Bid} and \textsc{Claim}) which are shown in bold in \cref{fig:contract_dispatcher}.
    
\begin{itemize}
    \item \textbf{Payable Check.} A \textsc{Solidity} function can be declared as \textit{payable} if it accepts transactions that have an associated \textsc{Ether} value.
    When a function is not declared as payable, the Ethereum compiler makes sure that the EVM reverts such transactions.
    Since our goal is to test only the functionalities implemented by the developer, \textsc{AGSolT} ignores such branches and simply does not send \textsc{Ether} to non-payable functions.
\end{itemize}

\begin{wrapfigure}{r}{0.5\textwidth}
    \centering
    \captionsetup{margin=0.03\textwidth}
    \includegraphics[width=0.45\textwidth]{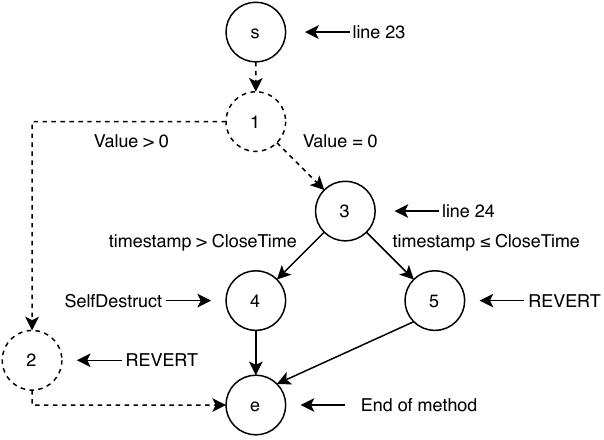}
    \caption{CDG of the {\tt Claim} function in Smart Contract \ref{sc:eth_specific}}
    %\Description[CDG of the {\tt Claim} function in Smart Contract \ref{sc:eth_specific}]{Control Dependency Graph of the {\tt Claim} function in Smart Contract \ref{sc:eth_specific}}
    \label{fig:claim}
\end{wrapfigure}

As an example, \Cref{fig:claim} shows the CDG of the {\tt Claim} function reported in Smart Contract \ref{sc:eth_specific}. Before going from line 23 to 24, the EVM verifies if the transaction has an \textsc{Ether} value and, if it does, reverts the transaction. \textsc{AGSolT} trims the dashed nodes and edges and merges the start node with node 3.

\subsection{Testing Loop}
\label{sec:testing_loop}

During the testing loop, the actual search for optimal test cases is performed until the budget is consumed. We first discuss the difference between a random- and a search-based algorithm, followed by the general steps.

\subsubsection{Random Initialisation of Test Cases}
\label{sec:test_cases_and_random_initialisation}

After extracting the required information, the population of test cases is initialized through a random algorithm.
As in previous work, each test case is a sequence of statements $t=\left<s_1, s_2, ..., s_n\right>$ ~\cite{Fraser2013,fraser2011mutation,tonella2004evolutionary}.
\textsc{AGSolT} relies on two types of statements:

\begin{itemize}
    \item \textbf{Constructor statements} are used to deploy smart contracts on the blockchain.
    Such statements are used as the first statement $s_1$ of each test case $t$ to ensure that a fresh instance of the smart contract is instantiated for each test case on which the function statements can be called.
    This statement type contains the information required to deploy an instance of the relevant smart contract on the blockchain, including the input variables required by the smart contract constructor and the transaction metadata, such as the amount of ETH send with the transaction and the account from which the transaction is sent.

    \item \textbf{Function statements} are used to create transactions that invoke functions in the deployed smart contracts.
    Indeed, the only way to interact with a smart contract in Ethereum is by sending a transaction to its address.
    All the statements, but the first (\ie the constructor statement), in a test case are function statements that are responsible for traversing the branches of the smart contract.
    This statement type contains a reference to the function to cover, its input variables, and the transaction metadata.
\end{itemize}

A set of test cases is initialized by creating $N$ random test cases, where $N$ is the population size \ie the number of test cases in any generation.
When \textsc{AGSolT} relies on the random search, test cases are generated by performing only this step.
The search keeps running through until either full branch coverage is achieved or the specified \textit{budget} is consumed.
At this point, the final population (\ie the archive) is presented as the solution.
As shown in \Cref{fig:Flowchart_AGSolT}, to improve the generated test cases, \textsc{AGSolT} can perform a guided search and a random search. For the former, we integrated \textsc{DynaMOSA} (\ie Many-Objective Sorting Algorithm with Dynamic target selection), the genetic algorithm proposed by Panichella \etal~\cite{panichella2018}.

\subsubsection{Genetic Loop: the DynaMOSA Algorithm}
\label{sec:dynamosa_alg}

Genetic Algorithms are inspired by biological evolution: they work with a population of (candidate) solutions or \textit{chromosomes} from which they derive a next \textit{generation} of solutions by iteratively applying \textit{evaluation}, \textit{selection}, \textit{crossover}, and \textit{mutation}.
Mitchell~\cite{mitchell1998} and Lucken \etal~\cite{Lucken2014} provide more details on genetic algorithms for multi-objective problems.
\textsc{DynaMOSA}~\cite{panichella2018} is a state-of-the-art algorithm specifically designed for automated test case generation.
It facilitates the creation of a small and effective test suite through multi-objective optimization inspired by NSGA-II~\cite{Deb-02}.
DynaMOSA has been shown to significantly outperform other test case generation algorithms (\eg Whole-Suite Approach~\cite{Fraser2013} and LIPS~\cite{Scalabrino2016,panichella2017lips}) in terms of branch and mutation coverage on an extensive set of Java classes.

\textbf{Fitness Function.} The search algorithm is guided by the normalized branch distance, as defined by Arcuri \etal~\cite{Fraser2013}.
The normalized branch distance for a test case $t$ and a branch $b$ with controlling predicate $p_b$ is given by:

\begin{equation}
\label{def:norm_bd}
    d(t,b)= 
\begin{cases}
    0 & \text{if $t$ satisfies $p_b$,}\\
    \frac{f_{p_b}(t,b)}{f_{p_b}(t, b)+1} & \begin{tabular}{@{}l@{}}\text{if $p_b$ has been reached but not } \\[-0.9ex] \text{satisfied,}\end{tabular} \\
    1 & \text{otherwise.}
\end{cases}
\end{equation}

Here $f_p(t,b)$ is given by Korel's objective function for relational predicates as shown in \Cref{tab:korel_functions}~\cite{korel90}.
Test cases with a smaller normalized branch distance are closer to covering the corresponding branch and are thus more desirable.

\begin{table}[!h]
    \centering
    \caption{Relational predicates and objective functions~\cite{korel90}}
    \begin{tabular}{cc}
        \hline
        Relational predicate & $f_p$ \\
        \hline
        $a>b$ & $b-a$\\
        $a\geq b$ & $b-a$\\
        $a<b$ & $a-b$\\
        $a\leq b$ & $a-b$\\
        $a=b$ & $abs(a-b)$\\
        $a\neq b$ & $-abs(a-b)$\\
        \hline
    \end{tabular}
    \label{tab:korel_functions}
\end{table}

Because we aim at covering all branches simultaneously, the goal of the search becomes the following, similarly to what previously formulated by Panichella \etal~\cite{panichella2018}:

\begin{dfn}[Fitness Function]
Let $B = \{b_1, b_2, ..., b_k\}$ be the set of branches in a smart contract.
Find a test suite $T =\{t_1, t_2, ..., t_n\}$ consisting of non-dominated test cases $t$ that simultaneously minimizes the fitness function for each branch $b\in B$, \ie minimizing the following $k$ objective functions:

\begin{equation}
\label{thm:many_objective_formulation}
    \begin{cases}
    f_1(t) & = al(t, b_1) + d(t, b_1)\\
    f_2(t) & = al(t, b_2) + d(t, b_2)\\
    \vdots & \\
    f_k(t) & = al(t, b_k) + d(t, b_k)\\
\end{cases}
\end{equation}

Here $al(t, b_i)$ is the \textit{approach level} of $t$ to $b_i$ (\ie the number of predicates between the closest branch executed by $t$ and $b_j$) and $d(b,t)$ is the minimal \textit{normalized branch} distance of $t$ to branch $b \in B$ as defined in \Cref{def:norm_bd}.
\end{dfn}

Note that in this multi-objective approach a distance is calculated for each objective (branch) so that rather than using a \textit{single} distance to describe the fitness of a test case $t$, a distance \textit{vector}, $\vec{d}_t = \left<d(t, b_1), d(t, b_2), ..., d(t, b_n)\right>$, is used.

\textbf{Selection Operation.}
After randomly initializing the first generation of test cases and measuring the branch distances, the test cases are ranked using their Pareto fronts~\cite{Deb-02} as the primary criterion.
Although in multi-objective optimization having solutions that make trade-offs between objectives is usually desirable, this is not the case for automated test case generation.
Indeed, only fully covered branches are relevant for the branch coverage, whereas a test that \textit{almost} covers one (or more) uncovered branches does not add any value to the final test suite. 
When ranking test cases in the first Pareto front, \textsc{DynaMOSA} uses a preference criterion that generalizes this idea by determining, for each branch $b\in B$, the non-dominated test cases closest to covering $b$ and (if there are more than one) the shortest one among those.
More formally, as defined by Panichella \etal~\cite{panichella2018}, the preference criterion is the following:

\begin{dfn}[Preference Criterion.]
\label{def:pref_criterion}
Given a branch $b_i$ with corresponding objective function $d_i=d(b_i, t)$, a test case $t$ is preferred over another test case $t'$ (written as $t\prec_{b_i}t'$) \textit{iff}
\begin{equation}
d_i(t) < d_i(t') \text{ OR } d_i(t) = d_i(t') \land size(t) < size(t').
\end{equation}

where $size$ is a function that gives the length (\eg number of statements) of a given test case.
Size is considered a secondary criterion to prioritize solutions because shorter solutions reduce the oracle cost for humans~\cite{baresi2010,Fraser2013}.
\end{dfn}

To compare the test cases that are not in the same Pareto front and are not preferred by the preference-criterion, the \textit{sub-vector-distance-assignment} algorithm introduced by K{\"o}ppen and Yoshida~\cite{Koppen2014} is used as a secondary selection criterion.
Its goal is selecting the most diverse possible subset of solutions from the last Pareto front for the next generation.

\textbf{Crossover and Mutation Operations.}
We apply crossover and mutation on a test case level, following the approach suggested by Arcuri \etal~\cite{arcuri2013}. First, two parent test cases $p_1$ and $p_2$ are selected from the previous generation using tournament selection~\cite{Deb-02}. Each parent is then cut into two parts, and the first part from $p_1$ is combined with the second part of $p_2$ and vice versa two create two child test cases.
Mutation is performed by randomly applying \textit{remove, change} and \textit{insert} operators. These operators remove statements, slightly change the variables in the statements or insert new statements into test cases, respectively.

\subsubsection{Deploy \& Run Test Cases}
\label{sec:deploy_and_run_test_cases}
Before evaluating and selecting the best test cases for the next generation, each test case runs on an Ethereum blockchain environment.
As mentioned in \Cref{sec:test_cases_and_random_initialisation}, each test case starts with a constructor statement, which is used to deploy a new instance of the smart contract to the blockchain instance.
By looking at the receipt of the transaction, \textsc{AGSolT} instantiates the new smart contract and extracts its address on the blockchain.
Afterward, each method call is executed by sending a transaction to the instance's address.
The hash-codes of the transactions, which identify each transaction on the blockchain, are stored for the next step.

\subsubsection{Extract Execution Logs}
\label{sec:extract_execution_logs}
To compute the branch coverage for a test case as defined in \Cref{def:norm_bd}, two types of information are required: (i) the parts of the code covered by the test case and (ii) the values that are on the stack when a branch-controlling predicate is evaluated.
\textsc{AGSolT} extracts this information through a slightly modified functionality of the javascript \textsc{web3}\textsuperscript{\ref{ftn:web3}} debug module called \textit{getTransactionTrace}.
This module takes the transaction of a method call, recreates the blockchain state when the transaction was executed, and writes the executed opcodes and the stack evolution in a file used for the next evaluations.

\begin{algorithm}
    \caption{Evaluate Test Case}
    \label{alg:evaluate_test_case}
    \footnotesize
    \begin{flushleft}
    \textbf{Input:}\\
    $methodCalls$
    \Comment{The list of methods called by the test case}\\
    $Opcodelists$ 
    \Comment{The lists of opcodes executed by each Methodcall}\\
    $Callstacklists$ 
    \Comment{The lists of all the items on the stack when each opcode was executed}\\
    $Edges = \{E_1, E_2, ..., E_n\}$
    \Comment{The (ordered) list of Edges of the smart contract.}\\
    $Nodes = \{N_1, N_2, ..., N_m\}$
    \Comment{The (ordered) list of Nodes of the smart contract.}\\
    \textbf{Result:} a distance vector which contains the test case's distance to each branch.
    \end{flushleft}
    \begin{algorithmic}[5]
    \Procedure{Set Distances}{}
    \State $test\_scores = [\infty, \infty, ..., \infty]$
    \Comment{Distance to each Edge}
    \State $traversed = \emptyset$
    \Comment{The set of traversed edges}
    \ForEach{$Methodcall, Opcodelist, Callstacklist$}
    \State $curNode = startNode$
    \While{$curNode \neq endNode$}
    \State $nextNode =$ {\sc FindNextNode}($curNode, Opcodelist$)
    \ForEach{$E_i \in Edges$}
    \If{$E_i.startNode == curNode$}
    \State $test\_scores[i] = \min(test\_scores[i],$
    \State {\sc BranchDist}$(Opcodelist, Callstacklist, E_i))$
    \EndIf
    \If{$E_i.endNode == nextNode$}
    \State $traversed = traversed\bigcup\{E_i\}$
    \EndIf
    \EndFor
    \State $curNode = nextNode$
    \EndWhile
    \EndFor
    \ForEach{$E_i \in Edges$}
    \If{$test\_scores[i] == \infty$}
    \State $test\_scores[i] =$ {\sc ApproachLevel}$(E_i, traversed)$
    \EndIf
    \EndFor
    \EndProcedure
    \end{algorithmic}
\end{algorithm}

\subsubsection{Evaluate Test Cases \& Update Archive}
\label{sec:evaluate_test_cases}
After executing all the test cases and retrieving the necessary information, test cases are evaluated, as shown in \Cref{alg:evaluate_test_case}, to produce the distance vector, $test\_scores$, describing the test case's fitness.
For each test case, its distance to all branches is initialized as infinite (line 2).
Additionally, \textsc{AGSolT} keeps track of all traversed edges (initialized at line 3) to calculate the approach levels for those edges whose starting nodes are not reached during the execution.
For every method call in the test case, \textsc{AGSolT} takes the corresponding list of executed opcodes and a list of lists containing all the values on the stack when executing each opcode (line 4).
The first node in the CDG of any method is always the same (line 5), while its end is only reached when a node has no outgoing edges (line 6).
Finding the next node using the \textsc{FindNextNode} (line 7) method means looking at the first opcodes executed after leaving the current node and comparing them to the opcodes of the nodes with an incoming edge from the current node.
For each reached node, \textsc{AGSolT} analyzes the outgoing edges (lines 8-9) and updates the $test\_scores$ if the normalized branch distance from \Cref{def:norm_bd} is smaller than the smallest distance found so far in the test case (lines 10-11).
After identifying all traversed edges, \textsc{AGSolT} calculates for each not covered branch, the approach level: \ie the number of edges that would need to be traversed before the node controlling the branch can be reached (lines 19-23).
Finally, if a test case outperforms the best test-case found so far for a particular branch, it is stored in an \textit{archive}, which keeps track of the best test-case for each branch.
It is important to note that (as can be seen in \cref{fig:Flowchart_AGSolT}) both the random testing approach and the genetic approach go through steps \ref{sec:deploy_and_run_test_cases} through \ref{sec:evaluate_test_cases}.
The key difference between these approaches is that genetic algorithms use \textit{selection}, \textit{crossover}, and \textit{mutation} to create the next generation of test cases, while random testing creates a new set of randomly initialized test cases.

\subsection{Dealing with Blockchain Properties}
\label{sec:dealing_with_blockchain_properties}
To deal with the challenges that arise from transaction properties, blockchain properties, and interactive properties mentioned in \Cref{sec:problem_statement}, \textsc{AGSolT} provides configuration options that deal with the Ethereum and Solidity blockchain and smart contract environment:

\begin{itemize}
    \item \textbf{Transaction Properties.}
    \textsc{AGSolT} extracts all the accounts of the blockchain environment and uses them both as senders of transactions and as input variables whenever an address type is required.
    It keeps track of whether a function is payable and, if so, it sends an amount (between a configurable maximum and minimum) of Ether with the transaction.
    Both addresses and values can be evolved by the genetic algorithm as though they were input variables.
    
    \item \textbf{Blockchain Properties.}
    \textsc{AGSolT} allows the user to include a \texttt{PassBlocks} or  \texttt{PassTime} method call in test cases, which instruct the blockchain environment to update the latest block number or the time rather than invoke smart contract functions (assuming the chosen blockchain environment allows these manipulations).
    Both block number and time can only \textit{increase}, similarly to real-world Ethereum implementations.
    The miner configurations can be set in the blockchain environment.
    Therefore, they are not manipulated in \textsc{AGSolT}.
    
    \item \textbf{Interactive Properties.}
    In addition to using the extracted accounts as input variables, \textsc{AGSolT} has an option to include specific non-existent accounts, which can trigger specific errors. This feature also allows the users to indicate a new contract creation through the zero-address (\textsc{0x0}).
    Smart contracts can be deployed in the blockchain environment before the test suite generation.
    Their address can be provided to \textsc{AGSolT} as address input variables to test the interaction between the contracts.This offers a simple, yet effective way, for users to create e.g., stubs with their own desired functionality and test interaction properties. 
    Additionally, the user can use this functionality to provide addresses that do not exist on the blockchain as input variables for the contract functions to test the behavior of the contracts when the sent transactions fail.
\end{itemize}

\subsection{Resulting Test Suites}
At the end of the procedure shown in \cref{fig:Flowchart_AGSolT} \textsc{AGSolT} outputs a text-file that gives information about the test suite and the test process.
In particular, it includes the number of branches found and covered, the number of iterations through the loop before stopping, the total time spent testing, and the time spent running the tests on the blockchain.
Afterward, the test cases are provided as construct statements and method calls with relevant input- and transaction arguments. The test suite is easily interpretable for humans and can easily be automatically transformed into input for the user's preferred testing environment.

In addition to the test suite, \textsc{AGSolT} writes out the same meta information of \textit{all} contracts that were tested in a CSV file for easy comparison.

\section{Empirical Evaluation}
\label{sec:experimental_evaluation}
\begin{table*}[!ht]
    \centering
    \caption{The smart contracts used for evaluating \textsc{AGSolT} and their characteristics. Comm. stands for the "communication" domain.}
    \resizebox{1\linewidth}{!}{
    \begin{tabular}{|l|l|c|c|c|c|c|c|c|c|c|c|}
        \hline
        \textbf{Contract Name} & \textbf{Domain} &  \begin{tabular}{@{}c@{}}\textbf{$\#$ State-} \\ \textbf{ments}\end{tabular} & \begin{tabular}{@{}c@{}}\textbf{$\#$ Bran-} \\ \textbf{ches}\end{tabular} &
        \textbf{Found} & 
        \begin{tabular}{@{}c@{}}\textbf{Sender} \\ \textbf{Dep.}\end{tabular} & 
        \begin{tabular}{@{}c@{}}\textbf{Value} \\ \textbf{Dep.}\end{tabular} & \begin{tabular}{@{}c@{}}\textbf{Acc. as} \\ \textbf{Vars}\end{tabular} &   \begin{tabular}{@{}c@{}}\textbf{NE Acc.} \\ \textbf{Dep.}\end{tabular} & \begin{tabular}{@{}c@{}}\textbf{Zero Acc.} \\ \textbf{Dep.}\end{tabular} &
        \begin{tabular}{@{}c@{}}\textbf{Block} \\ \textbf{Dep.}\end{tabular} & \begin{tabular}{@{}c@{}}\textbf{Time} \\ \textbf{Dep.}\end{tabular} \\
        \hline
        AddressBook & Comm. & 19 & 54 & \xmark & \cmark & \xmark & \cmark & \xmark & \xmark & \xmark & \xmark\\
        array-utils & Storage & 144 & 257 & \xmark & \xmark & \cmark & \xmark & \xmark & \xmark & \xmark & \xmark\\
        BadAuction & Token & 7 & 7 & \xmark & \cmark & \cmark & \xmark & \xmark & \xmark & \xmark & \xmark\\
        BasicToken & Token & 11 & 8 & \xmark & \cmark & \cmark & \cmark & \cmark & \xmark & \xmark & \xmark\\
        Casino & Exploit & 38 & 29 & \xmark & \cmark & \cmark & \xmark & \xmark & \xmark & \xmark & \cmark\\
        DateTime & Time & 90 & 143 & \xmark & \xmark & \xmark & \xmark & \xmark & \xmark & \xmark & \xmark\\
        DosAuction & Exploit & 7 & 7 & \cmark & \cmark & \cmark & \xmark & \xmark & \xmark & \xmark & \xmark\\
        \begin{tabular}{@{}l@{}}EIP20Standard-\\Token\end{tabular} & Token & 24 & 13 & \cmark & \cmark & \cmark & \cmark & \xmark & \xmark & \xmark & \xmark\\
        \begin{tabular}{@{}l@{}}EasyPayAnd-\\WithDraw\end{tabular} & Token & 7 & 8 & \xmark & \cmark & \cmark & \xmark & \xmark & \xmark & \xmark & \xmark\\
        EtherBank & Exploit & 13 & 17 & \xmark & \cmark & \cmark & \cmark & \xmark & \xmark & \xmark & \xmark\\
        EzToken & Token & 31 & 11 & \cmark & \cmark & \cmark & \cmark & \xmark & \xmark & \xmark & \xmark\\
        FixedSupplyToken & Token & 39 & 22 & \cmark & \cmark & \cmark & \cmark & \xmark & \xmark & \xmark & \xmark\\
        FundRaising & Finance & 23 & 21 & \xmark & \cmark & \cmark & \xmark & \xmark & \xmark & \xmark & \cmark\\
        Gift\_1\_ETH & Exploit & 18 & 18 & \cmark & \xmark & \cmark & \xmark & \xmark & \xmark & \xmark & \xmark\\
        Greeter & Comm. & 15 & 81 & \xmark & \xmark & \xmark & \xmark & \xmark & \xmark & \xmark & \xmark\\
        Greeter2 & Comm. & 13 & 60 & \xmark & \xmark & \xmark & \xmark & \xmark & \xmark & \xmark & \xmark\\
        Greeter3 & Comm. & 15 & 73 & \xmark & \xmark & \xmark & \xmark & \xmark & \xmark & \xmark & \xmark\\
        GuardCheck & Finance & 10 & 14 & \xmark & \cmark & \cmark & \cmark & \xmark & \cmark & \xmark & \xmark\\
        \begin{tabular}{@{}l@{}}GuessTheNum-\\berChallenge\end{tabular} & Exploit & 6 & 8 & \xmark & \xmark & \cmark & \xmark & \xmark & \xmark & \xmark & \xmark\\
        Identity & Identity & 53 & 131 & \xmark & \cmark & \cmark & \xmark & \xmark & \xmark & \xmark & \xmark\\
        IdentityManager & Identity & 49 & 90 & \cmark & \cmark & \xmark & \cmark & \xmark & \xmark & \xmark & \xmark\\
        LotteryFor10 & Betting & 45 & 44 & \cmark & \cmark & \cmark & \xmark & \xmark & \xmark & \cmark & \xmark\\
        \begin{tabular}{@{}l@{}}LotteryMultiple-\\Winners\end{tabular} & Betting & 31 & 45 & \xmark & \cmark & \cmark & \xmark & \xmark & \xmark & \xmark & \xmark\\
        MultiSigWallet (1) & Wallet & 56 & 70 & \xmark & \cmark & \cmark & \cmark & \cmark & \xmark & \xmark & \xmark\\
        MultiSigWallet (2) & Wallet & 59 & 83 & \xmark & \cmark & \cmark & \cmark & \cmark & \xmark & \xmark & \xmark\\
        MyAdvancedToken & Token & 53 & 3 & \cmark & \cmark & \cmark & \cmark & \xmark & \xmark & \xmark & \xmark\\
        OpenAddressLottery & Betting & 30 & 34 & \cmark & \cmark & \cmark & \cmark & \xmark & \xmark & \xmark & \xmark\\
        PermissionGroups & Identity & 58 & 86 & \cmark & \cmark & \xmark & \cmark & \xmark & \cmark & \xmark & \xmark\\
        Prover & Comm. & 27 & 17 & \cmark & \cmark & \xmark & \cmark & \xmark & \xmark & \xmark & \xmark\\
        Randomness & Betting & 22 & 17 & \xmark & \cmark & \xmark & \xmark & \xmark & \xmark & \xmark & \xmark\\
        Reentrance & Exploit & 9 & 14 & \cmark & \cmark & \cmark & \cmark & \xmark & \xmark & \xmark & \xmark\\
        Rubixi & Exploit & 56 & 102 & \cmark & \cmark & \cmark & \cmark & \xmark & \xmark & \xmark & \xmark\\
        SecureAuction & Finance & 11 & 6 & \cmark & \cmark & \cmark & \xmark & \xmark & \xmark & \xmark & \xmark\\
        TestDateTime & Time & 160 & 252 & \cmark & \xmark & \xmark & \xmark & \xmark & \xmark & \xmark & \xmark\\
        theRun & Exploit & 62 & 83 & \cmark & \cmark & \cmark & \cmark & \cmark & \xmark & \xmark & \xmark\\
        VulnerableTwoStep & Exploit & 11 & 10 & \xmark & \cmark & \cmark & \xmark & \xmark & \xmark & \xmark & \xmark\\
        \hline
        \end{tabular}}
    \label{tab:data_description}
\end{table*}

This section reports the empirical study that we performed to compare \textit{effectiveness}, \textit{efficiency}, and \textit{test case length} of the two algorithms for test case generation implemented in \textsc{AGSolT}: namely, a \textit{fuzzer} and \textit{DynaMOSA}~\cite{panichella2018}.

\subsection{Data Collection}
\label{sec:data_collection}
In an attempt to test on real-world smart contracts for our experiment, we scraped Github to obtain the most starred projects containing Solidity files. We selected the smart contracts that adhered to the following criteria: (i) being stand-alone, meaning they do not call other smart contracts during run-time (although they can inherit functionality from other smart contracts), (ii) coming from different application domains, (iii) not having any user-defined inputs for their functions.
We retrieved 36 Solidity smart contracts from 17 different repositories, which is comparable to existing studies ~\cite{Zhang2020,Liu2020MAM}.
To confirm that the contracts were used in the real world, we manually inspected them, and we found that at least 17 of the smart contracts have also been deployed on either the main Ethereum network or on a test network.
\cref{tab:data_description} shows the characteristics of the identified smart contracts, including their domain, whether they were found online, their number of statements, and number of branches in its CDG.
Additionally, \cref{tab:data_description} highlights presence the blockchain-specific qualities that \textsc{AGSolT} can handle. The \textit{sender dependence} and \textit{value dependence} indicate whether functionality of the smart contract depends on the transaction sender and transaction value and fall into the transaction properties discussed in \cref{sec:problem_statement} and \cref{sec:dealing_with_blockchain_properties}. \textit{Block dependence} and \textit{time dependence} indicate whether the contract relies on block number or the blockchain time for its functionality, which falls into the blockchain properties discussed in \cref{sec:problem_statement} and \cref{sec:dealing_with_blockchain_properties}. Finally \textit{account as variables}, \textit{non-existing account dependence} and \textit{zero account dependence} indicate the presence of interaction within the smart contract that would depend on the accounts passed as input variables and fall into the interaction properties discussed in \cref{sec:problem_statement} and \cref{sec:dealing_with_blockchain_properties}.
The entire data set, including the addresses of the deployed smart contracts, along with the tool and the results, is available in our online appendix\textsuperscript{\ref{ftn:agsolt_appendix}}.
The smart contracts are spread out over ten application domains. They vary in terms of the number of source code statements and branches in the CDG of the corresponding bytecode. We found that the transaction properties we identified occurred most frequently (28 sender dependencies and 26 value dependencies), followed by the interaction properties (29 variable dependencies, four non-existent account dependencies, and two zero-account dependencies). Interestingly only three smart contracts exhibited blockchain properties (two time dependencies and one block dependency).
This characteristic is feasible because relying on block and time information is inconsistent (each miner might have different information), and developers should rely on it as little as possible. Importantly only four smart contracts do not rely on any of the properties we identified. Since the presence of these dependencies was not part of the search protocol, this demonstrates the necessity for our tool (and others like it) to consider the blockchain-specific properties identified in \cref{sec:problem_statement}.

\subsection{\textsc{AGSolT} Evaluation}
To evaluate the effectiveness of \textsc{AGSolT} as well as compare the effectiveness of our random search and guided search, we perform an empirical study steered by the following research questions.

\begin{itemize}
    \item \textbf{RQ1 (Effectiveness).} \textit{Which is the coverage of the genetic algorithm approach compared to the random approach when generating test cases for Solidity smart contracts?}
    \item \textbf{RQ2 (Efficiency).} \textit{Which is the execution time of the genetic algorithm approach compared to the random approach when generating test cases for Solidity smart contracts?}
    \item \textbf{RQ3 (Test Case Length).} \textit{Which is the average number of statements in a test case for the genetic algorithm approach compared to the random approach when generating test cases for Solidity smart contracts?} 
\end{itemize}

The first two research questions are selected because they give insight into the performance of the two approaches as well as the general performance of \textsc{AGSolT}. The third research question is included because creating small, ``\textit{human-readable}'' test cases is a secondary objective of \textsc{DynaMOSA}~\cite{panichella2018}.

To answer the research questions, we implement both the random search and \textsc{DynaMOSA}-based guided search that were described in sections \ref{sec:test_cases_and_random_initialisation} and \ref{sec:dynamosa_alg} and run \textsc{AGSolT} for each approach and for each smart contract in \Cref{tab:data_description} to generate a test suite until either $i)$ full branch coverage is achieved or $ii)$ the tool has gone back to the start of the search loop in \Cref{fig:Flowchart_AGSolT} 100 times.
We repeated the process ten times for each smart contract to account for the inherent randomness of both approaches.
Our parameter settings for the genetic algorithm are the same as those used for evaluating \textsc{DynaMOSA}~\cite{panichella2018}, and the configurable options discussed in \Cref{sec:dealing_with_blockchain_properties} were appropriately set whenever possible to constrain the search.
In order to fairly compare the approaches and keeping with the above settings, we set the population size to 50 individuals for both approaches; therefore, the search budget consists of 5,000 test case evaluations or up to 200,000 method evaluations per smart contract.
As previously mentioned, we used \textsc{Ganache} to simulate the Ethereum blockchain, as it is much faster than a decentralized blockchain implementation. The execution was run on virtual machines running Ubuntu server with a RAM of 16GB.
For each generated test case, we measure its branch coverage, the time spent running tests on the blockchain, the total time, and the number of statements.
Additionally, we compute the statistical significance of the difference between the two approaches using \textit{Wilcoxon's test}~\cite{conover1980practical} with a p-value threshold of 0.05 as well as the Vargha-Delaney statistic ($\hat{A}_{12}$)~\cite{vargha2000critique} which is used to measure the magnitude of the difference.

\begin{table*}[ht]
    \centering
    \caption{Comparison for the achieved branch coverage for the genetic search algorithm and the fuzzing algorithm.}
    \begin{tabular}{|l|c|cc|cc|c|c|r|}
        \hline
        \multirow{2}{*}{\textbf{Name}} & \multirow{2}{*}{\textbf{$\#$ Branches}} & \multicolumn{2}{c|}{\textbf{Mean Cov. Gen.}} & \multicolumn{2}{c|}{\textbf{Mean Cov. Fuz.}} & \textbf{p-val} & \textbf{$\hat{A}_{12}$} & \textbf{Effect Size}\\
         & & \textbf{$\#$} & \textbf{\%} & \textbf{$\#$} & \textbf{\%} & & & \\
        \hline
        AddressBook & 54 & 54.0 & 1.00 & 54.0 & 1.00 & 1.00 & 0.50 & negligible\\
        BadAuction & 7 & 7.00 & 1.00 & 7.00 & 1.00 & 1.00 & 0.50 & negligible\\
        BasicToken & 8 & 8.00 & 1.00 & 8.00 & 1.00 & 1.00 & 0.50 & negligible\\
        Casino & 29 & 25.0 & 0.86 & 24.1 & 0.83 & \textbf{0.03} & \textbf{0.75} & large\\
        DosAuction & 7 & 7.00 & 1.00 & 7.00 & 1.00 & 1.00 & 0.50 & negligible\\
        EIP20StandardToken & 13 & 13.0 & 1.00 & 13.0 & 1.00 & 1.00 & 0.50 & negligible\\
        EasyPayAndWithDraw & 8 & 8.0 & 1.00 & 6.00 & 0.75 & \textbf{0.00} & \textbf{1.00} & large\\
        EtherBank & 17 & 14.0 & 0.82 & 14.0 & 0.82 & 1.00 & 0.50 & negligible\\
        EzToken & 11 & 11.0 & 1.00 & 11.0 & 1.00 & 1.00 & 0.50 & negligible\\
        FixedSupplyToken & 22 & 21.7 & 0.99 & 22.0 & 1.00 & 0.08 & 0.35 & small\\
        FundRaising & 21 & 21.0 & 1.00 & 21.0 & 1.00 & 1.00 & 0.50 & negligible\\
        Gift\_1\_ETH & 18 & 14.0 & 0.78 & 14.0 & 0.78 & 1.00 & 0.50 & negligible\\
        Greeter & 81 & 81.0 & 1.00 & 81.0 & 1.00 & 1.00 & 0.50 & negligible\\
        Greeter2 & 60 & 60.0 & 1.00 & 60.0 & 1.00 & 1.00 & 0.50 & negligible\\
        Greeter3 & 73 & 73.0 & 1.00 & 73.0 & 1.00 & 1.00 & 0.50 & negligible\\
        GuardCheck & 14 & 14.0 & 1.00 & 14.0 & 1.00 & 1.00 & 0.50 & negligible\\
        GuessTheNumberChallenge & 8 & 8.00 & 1.00 & 8.00 & 1.00 & 1.00 & 0.50 & negligible\\
        IdentityManager & 90 & 73.6 & 0.82 & 55.0 & 0.61 & \textbf{0.00} & \textbf{1.00} & large\\
        LotteryFor10 & 44 & 43.0 & 0.98 & 43.0 & 0.98 & 1.00 & 0.50 & negligible\\
        LotteryMultipleWinners & 45 & 44.7 & 0.99 & 43.4 & 0.96 & \textbf{0.05} & \textbf{0.78} & large\\
        MultiSigWallet (1) & 70 & 62.0 & 0.89 & 62.7 & 0.90 & 0.44 & 0.33 & medium\\
        MultiSigWallet (2) & 83 & 76.2 & 0.92 & 74.5 & 0.90 & 0.33 & 0.69 & medium\\
        MyAdvancedToken & 3 & 3.00 & 1.00 & 3.00 & 1.00 & 1.00 & 0.50 & negligible\\
        OpenAddressLottery & 34 & 32.0 & 0.94 & 32.0 & 0.94 & 1.00 & 0.50 & negligible\\
        PermissionGroups & 86 & 85.7 & 0.997 & 83.7 & 0.97 & \textbf{0.01} & \textbf{0.96} & large\\
        Prover & 17 & 17.0 & 1.00 & 17.0 & 1.00 & 1.00 & 0.50 & negligible\\
        Randomness & 17 & 16.0 & 0.94 & 16.0 & 0.94 & 1.00 & 0.50 & negligible\\
        Reentrance & 14 & 13.0 & 0.93 & 13.0 & 0.93 & 1.00 & 0.50 & negligible\\
        Rubixi & 102 & 67.0 & 0.66 & 69.0 & 0.68 & \textbf{0.00} & \textbf{0.05} & large\\
        SecureAuction & 6 & 6.00 & 1.00 & 6.00 & 1.00 & 1.00 & 0.50 & negligible\\
        TestDateTime & 252 & 243 & 0.96 & 240 & 0.95 & \textbf{0.02} & \textbf{0.75} & large\\
        theRun & 83 & 34.0 & 0.41 & 34.0 & 0.41 & 1.00 & 0.50 & negligible\\
        VulnerableTwoStep & 10 & 10.0 & 1.00 & 10.0 & 1.00 & 1.00 & 0.50 & negligible\\
        \hline
    \end{tabular}
    \label{tab:coverage_results}
\end{table*}

\begin{table}[ht]
    \centering
    \caption{Frequency of full branch coverage for the two approaches.}
    % \resizebox{1\linewidth}{!}{
    \begin{tabular}{|l|c|c|}
    \hline
    \textbf{Name} & \textbf{Full Cov. Gen.} & \textbf{Full Cov. Fuz.}\\
        \hline
                  AddressBook &                  10 &                  10 \\
                   BadAuction &                  10 &                  10 \\
                   BasicToken &                  10 &                  10 \\
                   DosAuction &                  10 &                  10 \\
           EIP20StandardToken &                  10 &                  10 \\
           EasyPayAndWithDraw &                  10 &                  -  \\
                      EzToken &                  10 &                  10 \\
             FixedSupplyToken &                   7 &                  10 \\
                  FundRaising &                  10 &                  10 \\
                      Greeter &                  10 &                  10 \\
                     Greeter2 &                  10 &                  10 \\
                     Greeter3 &                  10 &                  10 \\
                   GuardCheck &                  10 &                  10 \\
      GuessTheNumberChallenge &                  10 &                  10 \\
       LotteryMultipleWinners &                   7 &                   2 \\
           MultiSigWallet (1) &                   1 &                   - \\
              MyAdvancedToken &                  10 &                  10 \\
             PermissionGroups &                   8 &                   - \\
                       Prover &                  10 &                  10 \\
                SecureAuction &                  10 &                  10 \\
            VulnerableTwoStep &                  10 &                  10 \\
            \hline
\end{tabular}%}
\label{tab:full_cov_count}
\end{table}

\begin{table*}[ht]
    \centering
    \caption{Comparison for the time spend on creating tests for the genetic search algorithm and the fuzzing algorithm.}
    \resizebox{1\linewidth}{!}{
    \begin{tabular}{|l|cc|cc|cc|cc|c|c|r|}
        \hline
        \multirow{2}{*}{\textbf{Name}} & \multicolumn{2}{c|}{\textbf{Generations}} & \multicolumn{2}{c|}{\textbf{Time/Generation}} & \multicolumn{2}{c|}{\textbf{Total Time (s)}} & \multicolumn{2}{c|}{\textbf{Chain Time (\%)}} & \multirow{2}{*}{\textbf{p-value}} & \multirow{2}{*}{\textbf{$\hat{A}_{12}$}} & \multirow{2}{*}{\textbf{Effect Size}}\\
         & \textbf{Gen.} & \textbf{Fuz.} & \textbf{Gen.} & \textbf{Fuz.} & \textbf{Gen.} & \textbf{Fuz.} & \textbf{Gen.} & \textbf{Fuz.} & & & \\
        \hline
        AddressBook & 3.50 & 1.90 & 122 & 277 & 428 & 527 & 0.72 & 0.71 & 0.28 & 0.40 & small\\
        BadAuction & 1.00 & 1.00 & 85.9 & 86.5 & 85.9 & 86.5 & 0.84 & 0.84 & 0.96 & 0.5 & negligible\\
        BasicToken & 1.00 & 1.00 & 92.7 & 95.2 & 92.7 & 95.2 & 0.76 & 0.77 & 0.39 & 0.39 & small\\
        Casino & 101 & 101 & 80.3 & 133 & 8115 & 13432 & 0.82 & 0.82 & \textbf{0.01} & \textbf{0.00} & large\\
        DosAuction & 1.00 & 1.00 & 69.2 & 70.9 & 69.2 & 70.9 & 0.85 & 0.85 & 0.58 & 0.43 & negligible\\
        EIP20StandardToken & 1.00 & 1.00 & 1001 & 110 & 101 & 110 & 0.76 & 0.75 & \textbf{0.01} & \textbf{0.18} & \textbf{large}\\
        EasyPayAndWithDraw & 3.50 & 101 & 122 & 78.1 & 425 & 7888 & 0.84 & 0.86 & \textbf{0.01} & \textbf{0.00} & large\\
        EtherBank & 101 & 101 & 67.8 & 19.5 & 6848 & 1970 & 0.84 & 0.88 & \textbf{0.01} & \textbf{1.00} & large\\
        EzToken & 1.00 & 1.00 & 137 & 146 & 137 & 146 & 0.72 & 0.73 & \textbf{0.05} & \textbf{0.21} & \textbf{large}\\
        FixedSupplyToken & 34.7 & 4.10 & 79.5 & 172 & 2758 & 704 & 0.79 & 0.77 & 0.33 & 0.60 & small\\
        FundRaising & 1.00 & 1.00 & 81.1 & 79.7 & 81.1 & 79.7 & 0.77 & 0.77 & 0.80 & 0.52 & negligible\\
        Gift\_1\_ETH & 101 & 101 & 66.4 & 93.1 & 67078 & 9399 & 0.83 & 0.82 & \textbf{0.01} & \textbf{0.00} & large\\
        Greeter & 2.70 & 1.30 & 172 & 162 & 463 & 210 & 0.70 & 0.69 & 0.33 & 0.63 & small\\
        Greeter2 & 1.00 & 1.20 & 183 & 182 & 183 & 218 & 0.68 & 0.69 & \textbf{0.03} & \textbf{0.28} & medium\\
        Greeter3 & 2.10 & 1.60 & 172 & 250 & 361 & 400 & 0.71 & 0.68 & 0.28 & 0.20 & large\\
        GuardCheck & 1.00 & 1.00 & 86.2 & 74.0 & 86.2 & 74.0 & 0.82 & 0.83 & \textbf{0.02} & \textbf{0.75} & large\\
        GuessTheNumberChallenge & 1.90 & 1.30 & 52.3 & 35.5 & 99.5 & 46.2 & 0.84 & 0.75 & 0.33 & 0.23 & large\\
        IdentityManager & 101 & 101 & 141 & 113 & 14196 & 11430 & 0.73 & 0.76 & 0.09 & 0.70 & medium\\
        LotteryFor10 & 101 & 101 & 149 & 113 & 15039 & 11389 & 0.73 & 0.79 & \textbf{0.01} & \textbf{1.00} & large\\
        LotteryMultipleWinners & 65.9 & 91.7 & 174 & 95.1 & 11472 & 8721 & 0.76 & 0.79 & 0.28 & 0.72 & medium\\
        MultiSigWallet (1) & 97.9 & 101 & 149 & 96.8 & 14600 & 9778 & 0.75 & 0.78 & \textbf{0.01} & \textbf{1.00} & large\\
        MultiSigWallet (2) & 101 & 101 & 190 & 96.6 & 19210 & 9760 & 0.74 & 0.79 & \textbf{0.01} & \textbf{1.00} & large\\
        MyAdvancedToken & 1.00 & 1.00 & 135 & 139 & 135 & 139 & 0.71 & 0.71 & 0.72 & 0.39 & small\\
        OpenAddressLottery & 101 & 101 & 245 & 101 & 24729 & 10159 & 0.79 & 0.81 & \textbf{0.01} & \textbf{1.00} & large\\
        PermissionGroups & 66.7 & 101 & 132 & 157 & 8827 & 15878 & 0.74 & 0.78 & \textbf{0.01} & \textbf{0.10} & large\\
        Prover & 1.00 & 1.00 & 200 & 193 & 200 & 193 & 0.71 & 0.69 & 0.28 & 0.62 & small\\
        Randomness & 101 & 101 & 86.3 & 86.5 & 8720 & 8740 & 0.82 & 0.83 & 0.80 & 0.51 & negligible\\
        Reentrance & 101 & 101 & 59.6 & 93.0 & 6022 & 9391 & 0.84 & 0.83 & \textbf{0.01} & \textbf{0.00} & large\\
        Rubixi & 101 & 101 & 53.4 & 121 & 5393 & 12257 & 0.70 & 0.71 & \textbf{0.01} & \textbf{0.00} & large\\
        SecureAuction & 1.00 & 1.00 & 90.3 & 87.0 & 90.3 & 87.0 & 0.81 & 0.81 & 0.09 & 0.63 & small\\
        TestDateTime & 101 & 101 & 211 & 345 & 21276 & 34882 & 0.62 & 0.62 & \textbf{0.01} & \textbf{0.11} & large\\
        theRun & 101 & 101 & 104 & 91.8 & 10538 & 9269 & 0.75 & 0.80 & \textbf{0.03} & \textbf{0.79} & large\\
        VulnerableTwoStep & 1.00 & 1.00 & 71.3 & 73 & 71.3 & 72.6 & 0.84 & 0.83 & 0.65 & 0.48 & negligible\\
        \hline
        \textbf{Mean} & 45.5 & 49.4 & 120 & 123 & 5683 & 5984 & 0.77 & 0.77 & - & - & -\\
        \hline
        \end{tabular}}
    \label{tab:time_results}
\end{table*}

\subsection{Analysis of the Results}
First, all tables miss the results for three contracts, which returned an error.
We found that invoking some functions of \textsc{DateTime} and \textsc{Identity} could cost more Gas than the block limit and that calling a function in \textsc{Identity} with \textsc{AGSolT} can produce an out of bounds error; therefore, we excluded them from the performance evaluation.
However, we included these smart contracts in \cref{tab:data_description} since they demonstrate the usefulness of \textsc{AGSolT} as a tool capable of detecting errors in popular real-world smart contracts.

\subsubsection{RQ 1. (Effectiveness)}
\Cref{tab:coverage_results} shows the mean branch coverage in terms of branches covered and the percentage of total branches covered for both \textsc{DynaMOSA}~\cite{panichella2018} and the fuzzer approach. Overall, both approaches achieved good branch coverage, \cref{tab:full_cov_count} shows that \textsc{DynaMOSA} managed to achieve full branch coverage for 21, while the fuzzer achieves full branch coverage for 18 smart contracts. Full branch coverage could not be achieved for several reasons. For example, some branches may be infeasible, or \textsc{AGSolT} settings should be tweaked further. For example, we noted that the ``LotteryFor10'' contract had one branch that was consistently not covered and found that this was because longer test cases (containing more than 40 statements) were necessary to cover this branch. One notable outlier on which both approaches perform poorly is the ``theRun'' contract, which relies on the block hash to simulate randomness, which is something that cannot be manipulated by \textsc{AGSolT}.

\cref{tab:coverage_results} also reports $p$-values from a Wilcoxon test as well as the $\hat{A}_{12}$ and effect size from a Vargha-Delaney test comparing the distributions of the achieved branch coverages (in percentages) by applying the genetic and fuzzing approach each ten times per smart contract.
Looking closer at the $p$-values and Varghay-Delaney statistic, we see that \textsc{DynaMOSA} achieves significantly higher coverage ($p \leq 0.05$) than the fuzzer in six cases, each with large effect size. In contrast, the fuzzer significantly outperformed \textsc{DynaMOSA} only once, also with large effect sizes. Additionally, when \textsc{DynaMOSA} outperforms the fuzzer, the average branch coverage increases between $3$\% and $25$\%, while the fuzzer only achieves a $2$\% (2 branches) increase. {\color{blue} We manually investigated those smart contracts for which the guided search could achieve full branch coverage, while the random search could not. In every case, we found that the branches not reached by the random search resulted from nested if-else statements and assertions.} This observation is in line with existing literature~\cite{Shamshiri2018,harman2010} that suggests that genetic algorithms could prove beneficial when compared to random testing approach for exercising deeper functionalities in code.\\

\noindent\fbox{
    \parbox{0.95\columnwidth}{
        \textbf{The genetic algorithm (\ie \textsc{DynaMOSA}) significantly outperformed the fuzzing algorithm when generating test cases for six Solidity smart contracts, whereas the opposite happened only once.}
    }
}

\subsubsection{RQ 2. (Efficiency)}
\cref{tab:time_results} shows the average number of generations (including the (first) random initialisation) for both approaches as well as the mean total time spend and the average time per generation. Additionally the Chain Time column, shows the average percentage of time that was spend running the tests on our blockchain implementation (as opposed to evaluating- and generating new test cases).
Interestingly, on average both approaches are more or less equally fast: with \textsc{DynaMOSA} taking $45.5$ generations on average compared to $52.4$ generations for the fuzzer and $5,683$ seconds to $5,984$ seconds for the fuzzer. This is surprising because the \textsc{DynaMOSA} algorithm follows the additional selection, crossover and mutation steps described in \cref{sec:AGSolT}. One possible explanation for this is the preference criterion \ref{def:pref_criterion}, which guides the search towards smaller test cases. Smaller test cases, in turn, take up less time; especially since \cref{tab:time_results} shows that most of the time in our experiments was used running the test cases on the blockchain.
In order to properly compare the results for the two implementations we performed Wilcoxon tests and Vargha-Delany tests comparing the distributions of average run times for the smart contracts for each approach, the results of which are shown in the final 3 columns of \cref{tab:time_results}.

There are ten smart contracts for which \textsc{DynaMOSA} significantly ($p \leq 0.05$) outperformed the fuzzer (9 with large and 1 medium effect size). The faster performance for ``EasyPayAndWithDraw'' and ``PermissionGroups'' can be attributed to the fact that for these smart contracts the genetic approach manages to regularly achieve branch coverage before the budget is consumed, whereas the fuzzer does not. For the other smart contracts we speculate that the preference criterion (\ref{def:pref_criterion}) in \textsc{DynaMOSA}, which guides the search to smaller test cases, saves time when running the tests in the blockchain environment and evaluating their performance as described in sections \ref{sec:deploy_and_run_test_cases} through \ref{sec:evaluate_test_cases}.
There are seven smart contracts for which the fuzzing approach significantly outperformed the genetic search (each with large effect size). For each of these, we see that the fuzzer, spends a smaller percentage off time \textbf{off}-chain compared to \textsc{DynaMOSA}. This makes as the fuzzer bypasses the (computationally intensive) \textit{selection}, \textit{crossover} and \textit{mutation} steps described in \cref{sec:dynamosa_alg}.\\

\noindent\fbox{
    \parbox{0.95\columnwidth}{
        \textbf{\textsc{DynaMOSA} was significantly faster than the fuzzing algorithm on ten smart contracts, whereas the opposite happened seven times.}
    }
}

\subsubsection{RQ 3. Test Case Length}
\cref{tab:length_results} shows the average test case length (in number of statements) for the final solution presented by both the genetic algorithm and the fuzzing algorithm. This solution is an archive, which stores for each branch to be covered, the shortest test case that covers it. Even though the creators of \textsc{DynaMOSA} cite the use of a preference criterion as a means for reducing the size of the test cases in the final test suite, in this experiment implementing an archive resulted in fairly similar results, at first glance, compared to \textsc{DynaMOSA} averaging 4.96 statements and the fuzzer averaging 5.03 statements.

To better compare the results of the two approaches a Wilcoxon test and a Vargha-Delany test were performed comparing the distributions of the average test case lengths of the final test suites for each smart contract. \textsc{DynaMOSA} produced significantly shorter ($p\leq0.05$ test cases for five smart contracts (four with large effect size and one with medium effect size), each of which it was also significantly faster for as shown in \cref{tab:time_results}. This supports the theory that the smaller test cases found by the guided search can lead to an increase in efficiency when compared to a random search. 
The fuzzing approach yielded significantly smaller test cases in the final test suite for 4 smart contracts. For the ``EasyPayAndWithDraw'' smart contract this can be explained by the guided search, which achieves full branch coverage fairly quickly, whereas the fuzzer consumes the full budget and thus has many more opportunities to generate smaller test cases. For the other smart contracts (two with large effect size and one with medium effect size) the improvement is very minor: ranging from 0.15 to 0.42 statements on average. 
If instead we look only at those smart contracts for which the fuzzing approach and the genetic approach complete in the same number of generations the average test case length for \textsc{DynaMOSA} becomes 3.70 statements and the average test case length for the Fuzzer becomes 3.96 which is slightly bigger.

\noindent\fbox{
    \parbox{0.95\columnwidth}{
        \textbf{The genetic algorithm (\ie \textsc{DynaMOSA}) produced significantly smaller test cases in the final test suites when compared to the fuzzing algorithm for five smart contracts. The fuzzing algorithm produced significantly shorter test cases in the final test suites when compared to \textsc{DynaMOSA} for 3 smart contracts.}
    }
}

\begin{table}[!t]
    \centering
    \caption{Comparison between the average test case length for the genetic search algorithm and the fuzzing algorithm.}
    % \resizebox{1\textwidth}{!}{
    \begin{tabular}{|l|c|c|c|c|r|}
        \hline
        \textbf{Name} & \textbf{Gen.} & \textbf{Fuz} & \textbf{p-value} & \textbf{$\hat{A}_{12}$} & \textbf{Effect Size}\\
        \hline
        AddressBook & 9.39 & 10.1 & 0.39 & 0.44 & negligible\\
        BadAuction & 2.94 & 3.09 & 0.34 & 0.45 & negligible\\
        BasicToken & 3.67 & 3.54 & 0.61 & 0.50 & negligible\\
        Casino & 5.88 & 9.1 0& \textbf{0.01} & \textbf{0.10} & \textbf{large}\\
        DosAuction & 3.66 & 3.79 & 0.84 & 0.46 & negligible\\
        EIP20StandardToken & 5.71 & 5.89 & 0.58 & 0.42 & small\\
        EasyPayAndWithDraw & 8.1 & 2.0 & \textbf{0.01} & \textbf{1.00} & \textbf{large}\\
        EtherBank & 2.37 & 2.66 & \textbf{0.04} &\textbf{ 0.12} & \textbf{large}\\
        EzToken & 4.47 & 4.66 & 0.61 & 0.41 & small\\
        FixedSupplyToken & 4.15 & 4.85 & 0.28 & 0.41 & small\\
        FundRaising & 6.55 & 6.61 & 0.88 & 0.48 & negligible\\
        Gift\_1\_ETH & 2.02 & 2.01 & 0.32 & 0.60 & small\\
        Greeter & 7.62 & 7.51 & 0.96 & 0.52 & negligible\\
        Greeter2 & 7.01 & 7.34 & 0.72 & 0.44 & negligible\\
        Greeter3 & 8.22 & 9.16 & 0.09 & 0.26 & large\\
        GuardCheck & 4.46 & 4.71 & 0.44 & 0.40 & small\\
        GuessTheNumberChallenge & 14.46 & 13.69 & 0.54 & 0.50 & negligible\\
        IdentityManager & 3.44 & 3.24 & 0.44 & 0.44 & negligible\\
        LotteryFor10 & 4.49 & 4.46 & 0.72 & 0.48 & negligible\\
        LotteryMultipleWinners & 8.33 & 7.15 & 0.07 & 0.71 & medium\\
        MultiSigWallet (1) & 3.88 & 6.59 & \textbf{0.01} & \textbf{0.10} & \textbf{large}\\
        MultiSigWallet (2) & 3.85 & 6.81 & \textbf{0.01} & \textbf{0.07} & \textbf{large}\\
        MyAdvancedToken & 2.70 & 2.93 & 0.51 & 0.42 & small\\
        OpenAddressLottery & 2.51 & 2.09 & \textbf{0.02} & \textbf{0.80} & \textbf{large}\\
        PermissionGroups & 7.16 & 6.61 & 0.44 & 0.58 & small\\
        Prover & 4.41 & 4.42 & 0.57 & 0.56 & negligible\\
        Randomness & 2.46 & 2.58 & 0.07 & 0.28 & medium\\
        Reentrance & 2.15 & 2.0 & \textbf{0.05} & \textbf{0.70} & \textbf{medium}\\
        Rubixi & 3.73 & 3.52 & 0.58 & 0.50 & negligible\\
        SecureAuction & 3.87 & 3.58 & 0.24 & 0.65 & small\\
        TestDateTime & 2.42 & 2.14 & \textbf{0.01} & \textbf{0.97} & \textbf{large}\\
        theRun & 2.1 & 2.17 & \textbf{0.05} & \textbf{0.32} & \textbf{medium}\\
        VulnerableTwoStep & 5.33 & 5.15 & 0.80 & 0.52 & negligible\\
        \hline
        \textbf{Mean} & 4.96 & 5.03 &  & & \\
        \hline
        \end{tabular}%}
    \label{tab:length_results}
\end{table}

\section{Threats to Validity}
\label{sec:threats}
In this section, we discuss the threats to the validity of our experiment.

\textbf{Construct Validity.}
We demonstrated that transaction properties, blockchain properties, and interactive properties are present in some of the most popular Solidity smart contracts on Github. Additionally, we showed the effectiveness and efficiency of \textsc{AGSolT} by comparing a search-based test approach with a random testing one in terms of branch coverage, execution time, and test case length.
Both approaches were implemented in the same tool (\ie \textsc{AGSolT}) and executed on the same hardware environment to make the comparison as fair as possible.
We acknowledge that implementation issues could negatively impact the final results. However, please consider that we strictly followed the definition of the algorithm provided by Panichella \etal~\cite{panichella2018} and that our implementation is publicly available to allow other researchers to replicate our study.

\textbf{Internal Validity.}
All the experiments were executed ten times to address the inherent randomness of both approaches.
Fine-tuning the parameters of the DynaMOSA algorithm~\cite{panichella2018} could also have affected the internal validity of the experiments; since setting these parameters is challenging~\cite{arcuri2013}, we used the default values suggested by the creators of the algorithm~\cite{panichella2018}.

\textbf{External Validity.}
We tested \textsc{AGSolT} on a set of real-world smart contracts from a wide variety of developers.
We also ensured that each basic variable type and arrays in Solidity were included in the data set.
Despite exhibiting each of the properties that are indicative of the identified blockchain-specific challenges it is still possible that our data set is not representative of Solidity smart contracts in general and in general our results depend on the assumption that this data set is representative. Future experimentation with a larger data set is desirable.
\textsc{AGSolT} cannot yet handle user-defined input variable types nor smart contracts that rely on previously deployed smart contracts for their initialization.
Adding this feature is part of our research agenda.
Our conclusions are derived from the results obtained only on one genetic algorithm, namely DynaMOSA~\cite{panichella2018}.
Our research agenda includes experimentation with a broader set of search algorithms.
We did not run the test cases in a distributed blockchain, but we relied on \textsc{Ganache}, a framework to run tests, execute commands, and inspect smart contracts.
However, please consider that the resulting test suites are presented conveniently and can be easily used in any test network (\eg Ropstein\footnote{\url{https://ethereum.org/en/developers/docs/networks/\#testnet-faucets}}).

\textbf{Conclusion Validity.}
The results were obtained by repeating the experiments enough times and adopting appropriate statistical tests to draw valid conclusions. 
Specifically, we used the Wilcoxon test~\cite{conover1980practical} to test the significance of the differences and the Vargha-Delaney statistic~\cite{vargha2000critique} to estimate the effect size of the observed differences.

\section{Conclusion}
\label{sec:conclusion}
This paper discussed the challenges that arise when applying automated test case generation in a blockchain environment, identifying three different categories: transaction properties, blockchain properties and interactive properties. We presented, explored and partially validated \textsc{AGSolT}, a tool that addresses these challenges and creates test suites that aim to achieve branch coverage for Solidity smart contract unit testing.

\textsc{AGSolT} works with both a random testing approach (\ie a fuzzer) and a guided-search approach (\ie the \textsc{DynaMOSA} genetic algorithm~\cite{panichella2018}).
We gathered a data set consisting of real-world smart contracts from GitHub.
We demonstrated that many of these contracts exhibit behaviors that align with the challenges we identified.
Additionally, we have shown the effectiveness and efficiency of \textsc{AGSolT} by achieving good branch coverage with both approaches. In doing so we presented the first comparison between a guided search and a random search in the domain of automated test case generation for smart contracts.
We found that the \textsc{DynaMOSA} algorithm outperformed our fuzzer for achieving branch coverage, but ascertained that neither approach is significantly faster or produces significantly smaller test cases for the final test suite. The fact that the fuzzer was not faster, despite not going through the extra steps of selection, cross-over and mutation, is interesting and deserves further investigation. We hypothesize that this could be due to the preference criterion of \textsc{DynaMOSA}, which should, in theory, result in less time spent on the execution and evaluation steps of the testing procedure.
Finally, remarkably, we have shown that three of the most prevalent smart contracts on Github, suffer from critical failures (crashes) that emerged during our tests, demonstrating the potential real-world value of \textsc{AGSolT}.

In our future agenda, we plan to extend our current baseline a larger set of commercial smart contracts. Moreover we intend to leverage the parameterization of our testing approach with more search algorithms, e.g., the neural machine transaltion-based approach of Tufano et. al.~\cite{Tufano2020}. Additionally, we will expand \textsc{AGSoLT} to test inter-contract dependencies with the final goal of creating test cases in blockchain environments when multiple smart contracts interact. 

% \bibliography{references}

\end{document}